# Problems with Probability in Everett's Interpretation of Quantum Mechanics

Casey Blood
Sarasota, FL
CaseyBlood@gmail.com

**Abstract**
In the many-worlds interpretations (MWIs) of Everett and others, if I am the observer, there are several versions of me but no version is singled out as the one corresponding to my perceptions. However, it can be shown that the probability law implies one version must be singled out. Thus MWIs do not provide a sufficient basis for probability. If we are to have an acceptable description of the physical universe, MWIs must be supplemented by some mechanism, such as hidden variables or collapse, which singles out one version of the observer as the perceiving version.

## 1. Introduction.

Many-worlds interpretations (MWIs), as originally proposed by Everett [1] and elaborated upon by others [2]-[13], are among the major interpretations of quantum mechanics. They are of interest because they seem to require no amendments or additions—such as hidden variables or collapse—to the highly successful mathematics of quantum mechanics. The question to be considered is whether such interpretations are acceptable. A key feature of MWIs is that there are many equally valid versions of the observer in these interpretations, but none is singled out as *the* perceiving version. We argue (1) that this implies the probability law cannot even be stated in MWIs; and (2) that if the probability law is to hold, one version must be singled out as the version corresponding to perception. Since no version is singled out in MWIs, we conclude that many-worlds interpretations are not capable of accounting for the probability law. Some singling out mechanism must be added to the pure quantum mechanics of MWIs to account for the Born rule.

## 2. Quantum Mechanics and Versions of the Observer.

The basic or pure quantum mechanics used in MWIs—no hidden variables, no collapse, no "sentient beings," just the state vectors with all their branches—is called QMA here. Suppose we perform a measurement on an atomic-level system, with state vector, $\sum_{i=1}^{n} a_i |i\rangle$. After the measurement, if *I* am the observer, the state vector of the system in QMA is

(1) $\quad \Psi = \sum_{i=1}^{n} a_i |I_i\rangle |A_i\rangle |i\rangle$

The $|A_i\rangle$ are the *n* versions of the apparatus that detect and record the *n* possible outcomes, and the $|I_i\rangle$ are the *n* versions of the observer that perceive the readings on the versions of the apparatus. Note that the time evolution of the state vector is deterministic in QMA; there is nothing probabilistic in the mathematics.

There are two interesting characteristics of Eq. (1) and QMA. The first is that there are *n* equally valid *versions* of the observer but no singular "the" observer; that is, no version is singled out as the one corresponding exclusively to *my* experiential perceptions. If "*I*" perceive one result, there are *n* – 1 other, equally valid "*I*s" perceiving the other results. The second point is that the *n* versions of the observer *simultaneously* perceive their respective outcomes. This immediately makes one wonder how probability of perception is to be introduced into this scheme because, when all outcomes are simultaneously perceived, there can be no probability of perceiving one, specific outcome.

### 3. The Probability Law.

It would seem to be straightforward to state the probability law. If I am the observer, then the law says:

**P1**. The probability that I will perceive outcome *i* is $|a_i|^2$.

This statement is certainly correct experientially, but it is not acceptable *within the framework of QMA*. "I will perceive" implies perception of a single outcome by a unique *I*. But in QMA, there is no unique *I* that perceives just one outcome; instead there are *n* equally valid versions of *I*. So we might try a second statement, which acknowledges that only the versions perceive:

**P2**. The probability of my perceptions corresponding to those of version $|I_i\rangle$ is $|a_i|^2$.

But because there is a version of *me* associated with *every* outcome, *my* perceptions on each run correspond to those of *every* version, not just one version $|I_i\rangle$. And all the *n* perceptions simultaneously occur on every run, so there can be no *probability* of perceiving a single, given outcome. Thus neither **P1** nor **P2** can be used as the statement of the probability law in QMA. A third possible statement is:

**P3**. The probability of perceiving outcome *i* is $|a_i|^2$.

But this dodges the issue of what it is that perceives, and that is not acceptable in this context. We might try:

**P4**. The probability of outcome *i* occurring is $|a_i|^2$.

But that won't do either because all outcomes occur on every run of the experiment.

The point is this: The probability law is about what is perceived. And the only entities that perceive in QMA-MWI are the *n* versions of the observer. So

*the probability law in QMA-MWI must be written solely in terms of the perceptions of the versions of the observer*

with no reference to "my" perceptions or the perceptions of "the" observer. But that seems patently impossible because every version of the observer perceives its respective outcome on every run with 100% certainty; there is nothing probabilistic about the perceptions of the versions.

It has been argued [3]-[6], [9], [11]-[13] that probability can arise in deterministic QMA through a "subjective" process (see appendix C). But this does not help in stating the probability law in terms of the *perceptions* of the versions.

The conclusion is that it is not possible to properly state the probability law in QMA. And if it is not possible to *state* it, it is surely not possible for the law to hold in QMA. This inability of QMA to account for the probability law implies QMA, by itself, is not a sufficient basis for describing reality. That in turn means MWIs, which are based solely on QMA (plus, in some treatments, the assumption that the probability law holds), cannot be valid interpretations.

## 4. Singling Out.

By looking at a particular case, we can see the problem with MWIs in more detail. Suppose we consider a two-state case

(2) $\Psi = a_1 |I_1\rangle |A_1\rangle |1\rangle + a_2 |I_2\rangle |A_2\rangle |2\rangle$

with $|a_1|^2 = .999, |a_2|^2 = .001$, and suppose we do 10 runs, with the observer perceiving the results of every run. There will be $2^{10} = 1,024$ possible outcomes and 1,024 versions of the observer, each equally valid. But we know experientially that my perceptions will (almost always) correspond to only one of them, the one with all 10 outcomes 1. That is, one version from among all the 1,024 versions is *singled out* as the one corresponding to my experiential perceptions.

But in QMA, not only is each version equally valid, but also *every version* of the observer *perceives its respective outcome* on every run of 10. No version is singled out as *the* perceived version in QMA. However, we know that one version *is* singled out in "perceptual reality." The inescapable conclusion is that there must be a coefficient-sensitive mechanism (collapse?, hidden variables?), outside QMA, that singles out one version as *the* perceived version. That is, QMA-MWI must be supplemented if one is to obtain a sufficient basis for explaining the probability law.

## 5. Conclusion.

The above arguments show that QMA cannot accommodate the probability law, even if one simply *assumes* the law holds instead of attempting to deduce it from within QMA. For the probability law to hold, there must be a random, coefficient-dependent process that singles out version *i* as the perceived version on a fraction $|a_i|^2$ of the runs of the experiment. The singling out process could be collapse [14] [15], or hidden variables [16]-[18], or something else (see version 2). But since there is no singling out mechanism in many-worlds interpretations, including Everett's, these interpretations cannot be valid.

## Appendix A.
## The Auxiliary-Experiment Reasoning.

A line of reasoning has been proposed [5]-[8], [10] which, it is claimed, shows that if equal coefficients imply equal probability of perception, then the probability law can be deduced from QMA alone. It is my opinion that this reasoning contains an unwarranted assumption (or two). To illustrate, we will apply it to our $|a_1|^2 = .999, |a_2|^2 = .001$ case. The state is

(3) $\quad \Psi = \sqrt{\dfrac{999}{1000}} |I_1\rangle|A_1\rangle|1\rangle + \sqrt{\dfrac{1}{1000}} |I_2\rangle|A_2\rangle|2\rangle$

When the apparatus registers state 1 in this reasoning, it causes an experiment to be done on an auxiliary system, with the state of the auxiliary system being

(4) $\quad \displaystyle\sum_{j=1}^{999} \sqrt{\dfrac{1}{999}} |j\rangle'$

When we put (4) into (3) and assume the apparatus and the observer perceive the auxiliary experiment results, then the state is

$$(5) \quad \Psi = \sum_{j=1}^{999} \sqrt{\frac{1}{1000}} |I_{1,j}\rangle |A_{1,j}\rangle |1\rangle |j\rangle' + \sqrt{\frac{1}{1000}} |I_2\rangle |A_2\rangle |2\rangle$$

That is, there are now 1,000 versions of the observer, each with the same coefficient, $\sqrt{.001}$.

Now one can assume, or attempt to derive under various assumptions, that in the equal-coefficient case, *my* perceptions correspond to each possible outcome on an equal fraction of the runs.

> [I do not believe this is an allowable assumption in QMA. When *every* outcome is perceived on every run, there can no probability of perceiving a *particular* outcome, even when the coefficients are equal. Outcome $i$ is perceived on *every* run, not just $1/n$ of the runs.]

Then under this assumption, outcome 2 will be perceived only 0.1% of the time (from the single last term) and outcome 1 will be perceived—by the 999 different versions of observer 1, *each in a different universe*—99.9% of the time.

In addition to the questionable nature of the equal-coefficient assumption, the problem with this argument is that the same conclusion is presumed to hold when the auxiliary experiment is *not* done. But in that case, one must assume one can replace the single state, $\sqrt{.999}|I_1\rangle$, by 999 states of the observer, $|I_{1,j}\rangle, j=1,...,999$, *each in a different universe* (because the equal-coefficient-implies-equal-probability-of-perception assumption "holds" only if each version is in a different universe). I don't believe there is any way to justify that assumption.

> [One might try to use the "physical" reasoning that it shouldn't make any difference whether the auxiliary experiments are done or not. But the resulting states of the versions of the observer are just too different—in one case the Hilbert space for the states of *I* has dimension 2, in the other, 1,000—to allow this physical reasoning. In addition, reasoning from the physical situation doesn't seem appropriate when one is trying to make a point about the mathematics.]

## Appendix B.
## The Almost-All Strategy.

A probability law is a statement about what happens when an experiment is repeated many times. So let us consider *N* runs of a two-state experiment, with *N* very large. Then the state after the *N* runs is

$$(6) \quad \Psi_N = \sum_{m=0}^{N} \sum_{\{i(m)\}} |I(\{i(m)\})\rangle |A(\{i(m)\})\rangle |\{i(m)\}\rangle a_1^m a_2^{N-m}$$

where $\{i(m)\}$ denotes the values of a particular set of outcomes $i_1, i_2, ..., i_N, (i_j = 1 \text{ or } 2)$ for which $m$ of the $I$'s are 1. Note that, for a given $m$, the second sum runs over $N!/m!(N-m)!$ states, and there are a total of $2^N$ versions of the observer.

**Case 1:** $a_1 = a_2$ so all the coefficients in the sum are equal. Then, because $m$ is near $N/2$ for almost all of these states (from the binomial coefficient), almost all versions of the observer will perceive $m$ near $N/2$. Now the reasoning is (1) that *my* perceptions correspond to one of the versions of the observer and (2) that the corresponding version is "chosen," without bias, from the $2^N$ versions of the observer. If these two assumptions are granted, then *my* perceptions at the end of the $N$ runs will almost always correspond to perceiving $m$ near $N/2$, thus agreeing with the probability law in this simple case.

But this is not an allowable line of reasoning in QMA. The concept of *my* perception in (1)—implying exclusive perception by *me* of a single outcome—does not exist because there is no singular *me*; *every* outcome is simultaneously perceived by a *version* of me. So even in this simple case, the barrenness of QMA precludes support for the probability law. Once again, there can be no *probability* of perception when *every outcome is simultaneously perceived* by its respective version of the observer.

**Case 2**: $|a_1| > |a_2|$. It is still true in this case that almost all versions of the observer perceive $m$ near $N/2$. But this does not lead to Born's rule so the MWI advocate is forced to change assumption (2) to: (2*) for unknown reasons, the version corresponding to *my* perceptions is randomly chosen with a bias proportional to the coefficient ($a_1^m a_2^{N-m}$) squared. But the objection in case 1—that there is no *my* perception—still holds. The assumption (2*) implies there is a *me* that the biased choice applies to, as if the *me* were a (biased) spotlight focusing on one version. But that is not correct in QMA; singling out one version does not occur there.

Because we are precluded by assumption (2*) from perceiving the vast majority of states—those with $m$ near $N/2$—it is clear that a selection or singling out process is necessary to obtain the Born rule in the $N$-run case. But QMA does not provide such a process. So the probability law implies QMA-MWI must be supplemented by a singling out process. (And the singling out mechanism must be specified to have a complete picture of what governs our universe.)

# Appendix C.
# Subjective Probability.

One can argue [3]-[6], [9], [11]-[13] that from the point of view of a single version, there is an uncertainty in the outcome. After the measurement by the apparatus is made, but before the versions of the observer look at the readings, each version is asked, "What state are you in?" Each version is indeed ignorant of which universe he is in, and so that opens the door a crack for introducing probability. But I don't see how this uncertainty leads to any probability relevant to the Born rule.

*Within QMA* (rather than in the actual scheme that governs our universe), when a version is asked the probability of being in state $i$, why isn't the answer of every version just $1/n$, rather than $|a_i|^2$? Why should it depend on the coefficients? (Note that in QMA, the versions of the observer have no direct knowledge of the coefficients or of $n$.)

Actually there is a probability although it has nothing to do with the Born rule; the probability of version $j$ perceiving outcome $i$ is $\delta_{ij}$.

For an odd illustration of our objection, suppose each face of a biased die is intelligent. The die is rolled. After the roll, each face has a different number painted on it, but the faces cannot perceive the numbers. Then the "subjective probability" strategy is like asking each face, "What is the number painted on you?" It introduces a probability of sorts, but it doesn't seem relevant to any observation (such as which one might come out on top when the die is rolled).

To pursue the illustration further, suppose the die is rolled in the middle of empty space so there is no top face. When the faces are allowed to see the numbers, every outcome is simultaneously perceived, but no version is singled out as the "top" version (analogous to the one that is actually perceived in quantum mechanics). So there can be no probability of perceiving the "top" version. This seems to be an accurate analogy to the QMA situation.